\documentclass[journal]{IEEEtranTIE}

\usepackage{cite}
\ifCLASSINFOpdf
   \usepackage[pdftex]{graphicx}
\else
   \usepackage[dvips]{graphicx}
\fi
\usepackage{amsmath}
\usepackage{amssymb} 
\usepackage[caption=false,font=footnotesize]{subfig} 
\usepackage{epstopdf} 
\usepackage{url}
\usepackage{color} 
\usepackage{soul} 
\usepackage{placeins} 
\usepackage{blindtext} 
\usepackage{cancel} 
\usepackage{capt-of} 
\usepackage{enumitem} 
\usepackage{tikz} 
\usetikzlibrary{shapes.geometric, arrows}

\DeclareSymbolFont{matha}{OML}{txmi}{m}{it}
\DeclareMathSymbol{\varv}{\mathord}{matha}{118}

\definecolor{color_guard}{rgb}{0.01, 0.50, 0.10}
\definecolor{color_flow}{rgb}{0.2, 0.2, 0.6}
\definecolor{color_jump}{rgb}{1.0, 0.0, 0.0}

\begin{document}

\title{\huge{Hybrid Dynamical Model for Reluctance Actuators Including Saturation, Hysteresis and Eddy Currents}}

\author{
	\vskip 1em
	Edgar Ramirez-Laboreo, \emph{Student Member}, \emph{IEEE}, Maurice G. L. Roes, \emph{Member}, \emph{IEEE} \\ and Carlos~Sagues, \emph{Senior Member}, \emph{IEEE}

	\thanks{
		
		This work was supported in part by the Ministerio de Econom\'ia y Competitividad, Gobierno de Espa\~na - EU, under project RTC-2014-1847-6, in part by the Ministerio de Educaci\'on, Cultura y Deporte, Gobierno de Espa\~na, under grant FPU14/04171, and in part by project DGA-T45\_17R/FSE.

		E. Ramirez-Laboreo and C. Sagues are with the Departamento de Informatica e Ingenieria de Sistemas (DIIS) and the Instituto de Investigacion en Ingenieria de Aragon (I3A), Universidad de Zaragoza, Zaragoza 50018, Spain (e-mail: ramirlab@unizar.es, csagues@unizar.es).
		
		M. G. L. Roes is with the Department of Electrical Engineering, Eindhoven University of Technology, 5600 MB Eindhoven, The Netherlands (e-mail: m.g.l.roes@tue.nl).	
	}
    \thanks{\textcolor{red}{This is the accepted version of the manuscript: E. Ramirez-Laboreo, M. G. L. Roes and C. Sagues, ``Hybrid Dynamical Model for Reluctance Actuators Including Saturation, Hysteresis, and Eddy Currents," in IEEE/ASME Transactions on Mechatronics, vol. 24, no. 3, pp. 1396-1406, June 2019, doi: 10.1109/TMECH.2019.2906755. \textbf{Please cite the publisher's version}. For the publisher's version and full citation details see: \protect\url{https://doi.org/10.1109/TMECH.2019.2906755}. 
}}

    \thanks{© 2019 IEEE.  Personal use of this material is permitted.  Permission from IEEE must be obtained for all other uses, in any current or future media, including reprinting/republishing this material for advertising or promotional purposes, creating new collective works, for resale or redistribution to servers or lists, or reuse of any copyrighted component of this work in other works.}
}

\maketitle
\thispagestyle{plain}
\pagestyle{plain}

\begin{abstract}
A novel hybrid dynamical model for single-coil, short-stroke reluctance actuators is presented in this paper. The model, which is partially based on the principles of magnetic equivalent circuits, includes the magnetic phenomena of hysteresis and saturation by means of the generalized Preisach model. In addition, the eddy currents induced in the iron core are also considered, and the flux fringing effect in the air is incorporated by using results from finite element simulations. An explicit solution of the dynamics without need of inverting the Preisach model is derived, and the hybrid automaton that results from combining the electromagnetic and motion equations is presented and discussed. Finally, an identification method to determine the model parameters is proposed and experimentally illustrated on a real actuator. The results are presented and the advantages of our modeling method are emphasized.
\end{abstract}


\IEEEpeerreviewmaketitle

\markboth{IEEE/ASME TRANSACTIONS ON MECHATRONICS}
{}

\section{Introduction}

Reluctance actuators are being increasingly used in several domains because of their force density, fast response and efficiency \cite{katalenic2016high}. These features cause this type of actuators to be the ideal solution for novel high precision devices, e.g., antivibration systems for vehicles \cite{bao2014optimization} or aeronautical applications \cite{enrici2016design}, and may even make them advantageous with respect to classical induction motors \cite{kimiabeigi2016high}. Additionally, reluctance actuators are also ideal for small, low-cost electromechanical devices because of their compactness, reduced mass and low dissipation. In particular, single-coil short-stroke reluctance actuators are the basis of several commercial devices such as solenoid valves or electromagnetic relays, a class of devices that has recently received significant research attention in search of soft-landing behavior \cite{mercorelli2012antisaturating,yang2013multiobjective,ramirez2017new}.

In contrast to Lorentz actuators, which are driven by a force that is proportional to the coil current, the force that produces the motion in reluctance actuators is proportional to the square of the magnetic flux and is highly dependent on the position of the armature. In short, this high nonlinearity makes it very difficult to predict the force and to control the motion of reluctance actuators, especially for single-coil devices which can only apply magnetic force in one direction. Given that the armature position cannot be measured in practice---at least not with affordable sensors---many of the proposed solutions \cite{Lee2015control,Zhao2016linear,STRAUBERGER2016206} rely on dynamical models and estimation algorithms to achieve the designed control strategy. However, there are several electromagnetic phenomena, e.g., magnetic saturation and hysteresis, eddy currents, and flux fringing and leakage, that are usually neglected in the models in spite of having a strong influence on the dynamic behavior of the device \cite{vrijsen2014prediction}.

Different modeling methods have been adopted in the literature. On the one hand, the electromagnetic dynamics have been modeled basically by means of two different approaches: analytical parametric models based on the magnetic equivalent circuit (MEC) theory, which reduce the complexity of the system at the benefit of faster simulations, or numerical solutions based on the finite element method (FEM), which in general produce more accurate results at the expense of longer simulation times. Regarding the MEC approach, see, for instance, \cite{fang2016novel}, where a permanent magnet actuator is modeled by means of three magnetic circuits, or \cite{ramirez2016new}, where the authors propose a parametric reluctance which is experimentally identified. Alternatively, FEM models have been used, e.g., to optimize the design of linear actuators~\cite{bao2014optimization} and motors~\cite{kimiabeigi2016high}, or to estimate the attractive force of a circuit breaker \cite{xu2016multiphysics}, and they have been also combined with analytical dynamical models using curve fitting methods \cite{Gaeta2013experimental}. There are even certain works that use the MEC approach to obtain an analytical reluctance which is then validated or corrected by means of FEM simulations \cite{yang2013multiobjective,guofo2011output}, and some papers comparing the results of both approaches \cite{wattiaux2011modelling}. Mention should also be made of semianalytical methods (see, e.g., \cite{gysen2010general} and references therein). On the other hand, the motion dynamics have been also solved through different approaches. For instance, a mechanical model based on the Euler-Bernoulli theory for beams is proposed to predict the motion of a relay in~\cite{jun2008dynamic}, where a Kelvin-Voigt viscoelastic model is also proposed for modeling the contact bounce. The beam theory is used in other works~\cite{wattiaux2011modelling}, but the most widespread approach is the use of spring-damper rigid-body models with one~\cite{lin2013design} or two~\cite{ramirez2016new} degrees of freedom.

It is noteworthy that, given the extensive literature in the field, there is no model including all the aforementioned dynamic phenomena at the same time. There are several works that consider magnetic saturation \cite{bao2014optimization,Zhao2016linear,ramirez2016new,BRAUN2017782}, but nevertheless only a few including hysteresis \cite{STRAUBERGER2016206,mackenzie2016real} or eddy currents \cite{xu2016multiphysics,BRAUN2017782}. Special mention should be made of \cite{vrijsen2014magnetic}, which serves as a basis for the present work and may be considered the most comprehensive model of reluctance actuators with regard to the electromagnetic dynamics. However, neither the motion equations nor the flux fringing effect were included in the model because the investigated actuator was designed to work at a specific static position.

In this paper, we present a novel hybrid dynamical model for short-stroke single-coil reluctance actuators. The model, which includes eddy currents, magnetic saturation and hysteresis, is based on principles of the MEC theory but relies also on FEM simulations to describe the reluctance of the air gap. Consequently, not only does it describe the electromagnetic phenomena that exist within the iron, but it also accurately models the flux fringing effect in the air. The hybrid behavior of the model is due both to the position constraints that exist in the mechanism and to the use of the Preisach model for describing the magnetic hysteresis and saturation in the core. This latter model, which has been selected for being the most common approach in the literature, exhibits hybrid behavior because it uses different equations according to the direction of the magnetic field intensity. 

The main purpose of the model is to be used in processes that require a great number of accurate transient simulations, e.g., the design and validation of new estimation and control algorithms, where an FEM model would be impractical for computational reasons. In this regard, the main contributions of the research are as follows:
\begin{enumerate}[nosep]
\item A fast but accurate dynamical model for reluctance actuators including hysteresis, saturation, eddy currents, flux fringing and the motion of the armature.
\item The use of the coil voltage, which is the standard output of power sources, as input of the dynamical model instead of the current.
\item An explicit dynamical solution of the Preisach model without need of model inversion.
\item The proposal of a new class of hybrid dynamical systems \cite{goebel2009hybrid} whose state consists not only of scalar variables, but also of sets of constants with varying cardinality.
\end{enumerate}

The paper is organized as follows. First, in Section~\ref{sec:preisach}, the Preisach model for magnetic hysteresis and saturation is described and its corresponding incremental permeability, which is later used in the model of the actuator, is obtained. Then, in Section~\ref{sec:hybrid}, both the explicit dynamic solution of the Preisach model and the hybrid automaton that models the entire actuator are presented and discussed. In Section~\ref{sec:experiments}, an experimental setup including a real actuator is described and then used to identify the unknown parameters and to validate the proposed model.
Finally, the conclusions of the research are presented in Section~\ref{sec:conclusions}.

\section{Hysteresis and Saturation Model}
\label{sec:preisach}

\subsection{Classical Preisach Model}
\label{subsec:cpm}

\begin{figure}[t]
\centering
\includegraphics[]{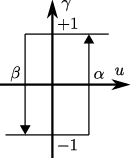}
\caption{Hysteron operator with threshold values $\alpha$ and $\beta$.}
\label{fig:hysteron}
\vspace{\floatsep}
\subfloat[]{\begin{minipage}[b][32mm][c]{37mm}\centering \includegraphics[]{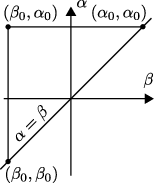} \end{minipage} \label{fig:preisach_plane_ab}}
\subfloat[]{\begin{minipage}[b][32mm][c]{47mm}\centering \includegraphics[]{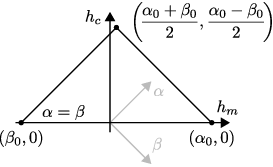} \end{minipage} \label{fig:preisach_plane_hmhc}}
\caption{Preisach plane in the (a) $\beta$--$\alpha$ and (b) $h_m$--$h_c$ coordinate systems.}
\label{fig:preisach_plane0}
\vspace{\floatsep}
\subfloat[]{\begin{minipage}[b][39mm][c]{42mm}\centering \includegraphics[]{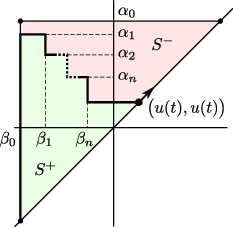} \end{minipage} \label{fig:preisach_plane_asc}}
\subfloat[]{\begin{minipage}[b][39mm][c]{42mm}\centering \includegraphics[]{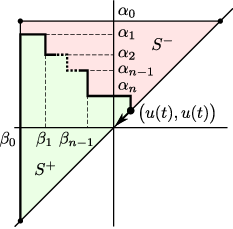} \end{minipage} \label{fig:preisach_plane_desc}}
\caption{Preisach plane division for (a) increasing input and (b) decreasing input. The input is assumed to start at its minimum possible value~$\beta_0$.}
\label{fig:preisach_plane_division}
\end{figure}

The classical Preisach model (CPM) \cite{preisach1935magnetische,mayergoyz1986mathematical} is based on an infinite set of basic hysteresis operators also known as \textit{hysterons} (see Fig.~\ref{fig:hysteron}). Each hysteron is characterized by two threshold values, $\alpha$ and $\beta$, which describe the output of the operator in terms of the time-dependent input $u={u}(t)$ as
\begin{equation}
\gamma\big(\alpha,\beta,{u},u_\mathrm{past}\big) = \left\{ \begin{array}{rcc}
+1, & \text{if} & {u}\geq\alpha \\
-1, & \text{if} & {u}\leq\beta \\
\gamma_\mathrm{past}, & \text{if} & \beta<{u}<\alpha
\end{array}
\right.
\end{equation}
where $u_\mathrm{past}$ is the last extremum of ${u}$ outside the interval $\left(\beta,\, \alpha\right)$ and $\gamma_\mathrm{past}$ is the hysteron output for that value. Considering $\alpha$ and $\beta$ as coordinates, the infinite hysterons are usually represented as points in the so-called Preisach plane (see Fig.~\ref{fig:preisach_plane0}). Since $\alpha$ is always greater than $\beta$, all of them are in fact located in the $\alpha>\beta$ half-plane.

The output of the CPM is then given by the sum of the outputs of the infinite number of hysteresis operators,
\begin{equation}
f_{\scriptscriptstyle\mathrm{CPM}}\left(u,\mathcal{U}\right) = \int_{\alpha>\beta} P(\alpha,\beta) \, \gamma\big(\alpha,\beta,{u},u_\mathrm{past}\big) \,\mathrm{d}\alpha \,\mathrm{d}\beta,
\label{eq:preisach_1}
\end{equation}
where $\mathcal{U}=\mathcal{U}(t)=\left\{\,u(\tau) \ | \ \tau<t \,\right\}$ contains the history of~$u$, from which $u_\mathrm{past}$ can be obtained for each hysteron, and $P(\alpha,\beta)$ is the Preisach function, which may be interpreted either as a weight function for an infinite set of homogeneously distributed hysterons or as a density function describing a non-homogeneous distribution of hysterons in the Preisach plane. Assuming that the input is bounded between two constants $\beta_0$ and $\alpha_0$, $\beta_0 \leq u(t) \leq \alpha_0$ $\forall\, t$, the Preisach function can be considered equal to zero outside the triangle with vertices $\left(\beta_0,\alpha_0\right)$, $\left(\alpha_0,\alpha_0\right)$, and $\left(\beta_0,\beta_0\right)$ (see Fig.~\ref{fig:preisach_plane_ab}).

In particular, due to the wiping-out property of the CPM \cite{mayergoyz1986mathematical}, the history of any arbitrary input $u(t)$ is fully characterized by a set of previous maxima, $\mathcal{A}=\mathcal{A}(t)$, and a set of previous minima, $\mathcal{B}=\mathcal{B}(t)$, such that
\begin{align}
\mathcal{A}(t) = \left\{\alpha_0\right\} \cup \left\{\, u(\tau) \mid \tau<t, \ \dot{u}(\tau)=0, \ \ddot{u}(\tau)<0, \right.& \nonumber\\
\mathrm{max}\,\mathcal{B}(\tau) < u(s) < u(\tau) \ \forall s \in (\tau,t) & \!\left. \right\}, \\
\mathcal{B}(t) = \left\{\beta_0\right\} \cup \left\{\, u(\tau) \mid \tau<t, \ \dot{u}(\tau)=0, \ \ddot{u}(\tau)>0, \right.& \nonumber\\
u(\tau) < u(s) < \mathrm{min}\,\mathcal{A}(\tau) \ \forall s \in (\tau,t) & \!\left. \right\}
\end{align}
That is, a maximum (minimum) of $u$ at time $\tau$ is part of $\mathcal{A}$ (\,$\mathcal{B}$\,) at time $t$ if and only if all $u$ after $\tau$ and before $t$ has been contained between that maximum (minimum) and the immediately preceding minimum (maximum). Assuming that ${u}$ is initialized from its minimum possible value, i.e., $u(-\infty)=\beta_0$, we have that $\left|\mathcal{B}\right| = \left|\mathcal{A}\right|$ for increasing input and $\left|\mathcal{B}\right| = \left|\mathcal{A}\right|-1$ for decreasing input, where the operator $\left|\,\cdot\,\right|$ denotes the cardinality of the set. We use the notation $\alpha_i$ and $\beta_j$ to refer respectively to the $i$th largest element of $\mathcal{A}\setminus\!\left\{\alpha_0\right\}$ and the $j$th smallest element of $\mathcal{B}\setminus\!\left\{\beta_0\right\}$. As shown in Fig.~\ref{fig:preisach_plane_division}, the previous sets divide the Preisach plane into two regions, $S^+(u,\mathcal{A},\mathcal{B})$ and $S^-(u,\mathcal{A},\mathcal{B})$, in which the output of the hysterons are respectively equal to $+1$ and $-1$. Hence, the integral \eqref{eq:preisach_1} can be partitioned into the positive and negative regions,
\begin{equation}
f_{\scriptscriptstyle\mathrm{CPM}}(u,\mathcal{A},\mathcal{B}) \!=\!\! \iint_{S^+}\!\!\!\!\! P(\alpha,\beta) \,\mathrm{d}\alpha \,\mathrm{d}\beta -\!\! \iint_{S^-}\!\!\!\!\! P(\alpha,\beta) \,\mathrm{d}\alpha \,\mathrm{d}\beta.
\label{eq:preisach_2}
\end{equation}
Additionally, defining the integral of the Preisach function over the triangle of vertices $\left(\beta_j,\alpha_i\right)$, $\left(\alpha_i,\alpha_i\right)$ and $\left(\beta_j,\beta_j\right)$ as
\begin{equation}
T(\alpha_i,\beta_j) = \int_{\beta_j}^{\alpha_i} \int_\beta^{\alpha_i}  P\left(\alpha,\beta\right) \,\mathrm{d}\alpha \, \mathrm{d}\beta,
\label{eq:preisach_triangle}
\end{equation}
the output of the CPM can be computed by addition and subtraction of triangle integrals. Hence, given $n=\left|\mathcal{A}\setminus\!\left\{\alpha_0\right\}\right|$, the output of the model for an increasing input (see Fig.~\ref{fig:preisach_plane_asc}) is given by
\begin{align}
{f_{\scriptscriptstyle\mathrm{CPM,incr}}}(u,\mathcal{A},\mathcal{B}) =& - T(\alpha_0,\beta_0) + 2\sum_{k=1}^n T(\alpha_k,\beta_{k-1})\nonumber \\[-1ex]
& - 2\sum_{k=1}^n T(\alpha_k,\beta_k) + 2\,T\big({u},\beta_n\big), \label{eq:yCPM_inc}
\end{align}
while, for a decreasing input (see Fig.~\ref{fig:preisach_plane_desc}), is equal to
\begin{align}
f_{\scriptscriptstyle\mathrm{CPM,decr}}(u,\mathcal{A},\mathcal{B}) =& - T(\alpha_0,\beta_0) + 2\sum_{k=1}^{n} T(\alpha_k,\beta_{k-1})\nonumber \\[-1ex]
& - 2\sum_{k=1}^{n-1} T(\alpha_k,\beta_k) -2\,T\big(\alpha_n,{u}\big). \label{eq:yCPM_dec}
\end{align}

Several analytical functions have been used in the literature as Preisach functions \cite{vrijsen2014magnetic}. The most general approach to construct such a function is to assume that $P$ can be expressed as the product of two univariate probability density functions depending on $h_c = (\alpha-\beta)/2$ and $h_m = (\alpha+\beta)/2$
(see Fig.~\ref{fig:preisach_plane_hmhc}), i.e.,
\begin{equation}
P(\alpha,\beta) = f_1\!\left(h_c\right) f_2\!\left(h_m\right) = f_1\!\left(\!\dfrac{\alpha-\beta}{2}\!\right) f_2\!\left(\!\dfrac{\alpha+\beta}{2}\!\right).
\end{equation}
In this case, the surface integral \eqref{eq:preisach_triangle} can be evaluated in the $h_m$--$h_c$ coordinate system and transformed analytically into a numerically less expensive line integral,
\begin{equation}
T(\alpha_i,\beta_j) = 2\int_0^{{\left(\alpha_i-\beta_j\right)}/{2}}{f_1(h_c) \Big[F_2(h_m)\Big]^{\alpha_i-h_c}_{\beta_j+h_c} \,\mathrm{d}h_c},
\label{eq:preisach_triangle_hmhc}
\end{equation}
where the factor of 2 is the Jacobian of the transformation from $\beta$--$\alpha$ to $h_m$--$h_c$ and $F_2(h_m)$ is the cumulative distribution function corresponding to $f_2(h_m)$.

The Cauchy distribution is known to be the best fit to most of the ferromagnetic materials~\cite{rouve1995application,azzerboni2003remarks,pruksanubal2006determination}. Its probability density and cumulative distribution functions are respectively given by
\begin{align}
f_{\scriptscriptstyle\mathrm{Cauchy}}(x\,|\,m_x,s_x) &= \dfrac{1}{\pi s_x\left(1+\left(\dfrac{x-m_x}{s_x}\right)^2\right)}, \\
F_{\scriptscriptstyle\mathrm{Cauchy}}(x\,|\,m_x,s_x) &= \dfrac{1}{2} + \dfrac{1}{\pi}\,\mathrm{arctan}\!\left(\dfrac{x-m_x}{s_x}\right),
\end{align}
where $m_x$ and $s_x$ are the parameters that specify the location and shape of the distribution. Consequently, in this work we approximate both $f_1$ and $f_2$ by means of this distribution:
\begin{align}
f_1\!\left(h_c\right) &= f_{\scriptscriptstyle\mathrm{Cauchy}}({h_c}\,|\,m_{h_c},s_{h_c}), \\
f_2\!\left(h_m\right) &= f_{\scriptscriptstyle\mathrm{Cauchy}}({h_m}\,|\,m_{h_m},s_{h_m}).
\end{align}
Note that, in order to obtain a symmetrical major hysteresis loop about the origin, $m_{h_m}$ must be equal to zero, $m_{h_m}\!=\!0$. Hence, the proposed Preisach distribution only depends on three parameters: $m_{h_c}$, $s_{h_c}$ and $s_{h_m}$.

\subsection{Limitations and Generalization}

When applied to modeling magnetic hysteresis, the CPM has two major drawbacks, namely the zero magnetic permeability in the reversal points and the non-saturating behavior of the magnetization $M$. These are solved in the so-called generalized Preisach model (GPM) \cite{mayergoyz1988generalized} by modeling the magnetic flux density $B$ as the sum of two terms depending on the magnetic field intensity $H$,
\begin{equation}
B = f_{\scriptscriptstyle\mathrm{GPM}}(H,\mathcal{A},\mathcal{B}) =  B_{\scriptscriptstyle\mathrm{Rev}}(H) + B_{\scriptscriptstyle\mathrm{Irr}}(H,\mathcal{A},\mathcal{B}),
\label{eq:BGPM}
\end{equation}
where $B_{\scriptscriptstyle\mathrm{Rev}}$ is the \textit{reversible} part, which is only dependent on the instantaneous value of $H$, and $B_{\scriptscriptstyle\mathrm{Irr}}$ is the \textit{irreversible} part, which depends also on the past maxima, $\mathcal{A}$, and minima, $\mathcal{B}$, of $H$. This latter part is obtained by means of the CPM,
\begin{equation}
B_{\scriptscriptstyle\mathrm{Irr}}(H,\mathcal{A},\mathcal{B}) = {\hat{B}_{\scriptscriptstyle\mathrm{Irr}}} \,\dfrac{f_{\scriptscriptstyle\mathrm{CPM}}(H,\mathcal{A},\mathcal{B})}{T(\alpha_0,\beta_0)},
\end{equation}
where ${\hat{B}_{\scriptscriptstyle\mathrm{Irr}}}$ is the saturation level of the irreversible part, $T(\alpha_0,\beta_0)$ acts as normalization factor, and $f_{\scriptscriptstyle\mathrm{CPM}}$ represents either $f_{\scriptscriptstyle\mathrm{CPM,incr}}$ or $f_{\scriptscriptstyle\mathrm{CPM,decr}}$ depending on the direction of $H$.

On the other hand, the reversible component, which provides the non-zero permeability at the reversal points, is commonly expressed as the integral with respect to $H$ of an incremental \textit{reversible} permeability,
\begin{equation}
B_{\scriptscriptstyle\mathrm{Rev}}(H) = \int_0^H \mu'_{\scriptscriptstyle\mathrm{Rev}}(H) \,\mathrm{d}H,
\label{eq:B_rev}
\end{equation}
where $\mu'_{\scriptscriptstyle\mathrm{Rev}}(H)$ is usually modeled by an analytical expression which is fitted to measurements. Based on the double exponential function used in \cite{vrijsen2014magnetic}, in this work we propose to model this permeability as 
\begin{equation}
\mu'_{\scriptscriptstyle\mathrm{Rev}}(H) = \mu_0 + \mu_1\,\mathrm{e}^{-\left|H\right|/H_1} + \mu_2\,\mathrm{e}^{-\left|H\right|/H_2},
\label{eq:mu_rev}
\end{equation}
where $\mu_0$ is the vacuum permeability and $\mu_1,\,\mu_2\in\mathbb{R}$ and $H_1,\,H_2\in\mathbb{R}^+$ are the parameters to identify. This expression leads to a reversible flux density equal to
\begin{align}
B_{\scriptscriptstyle\mathrm{Rev}}(H) = & \,\mu_0 H + \mathrm{sgn}(H)\,\mu_1 H_1\left(1-\mathrm{e}^{-\left|H\right|/H_1}\right) \nonumber\\
&  + \mathrm{sgn}(H)\,\mu_2 H_2\left(1-\mathrm{e}^{-\left|H\right|/H_2}\right).
\end{align}
Note that, while the use of the absolute value provides symmetry about the origin, the addition of the vacuum permeability in \eqref{eq:mu_rev} is what guarantees that $M$ saturates. In this regard, considering that the magnetization modeled by the GPM is $M = \big(f_{\scriptscriptstyle\mathrm{GPM}}(H,\mathcal{A},\mathcal{B}) /\mu_0-H\big)$ and that $\lim_{H\to\infty}B_{\scriptscriptstyle\mathrm{Irr}} = \hat{B}_{\scriptscriptstyle\mathrm{Irr}}$, the saturation level is easily obtained as
\begin{equation}
\lim_{H\to\infty}M =\dfrac{\mu_1 H_1 + \mu_2 H_2 + \hat{B}_{\scriptscriptstyle\mathrm{Irr}}}{\mu_0}.
\end{equation}
This can be used, for instance, to set physically meaningful initial values for the parameters to identify.

\subsection{Time Derivative of the GPM}
\label{subsec:mu_inc}

In order to use the GPM in a dynamical model, the time derivative of $B$ is obtained from \eqref{eq:BGPM} as
\begin{equation}
\dot{B} =  \left( \dfrac{\partial B_{\scriptscriptstyle\mathrm{Rev}}}{\partial H} + \dfrac{\partial B_{\scriptscriptstyle\mathrm{Irr}}}{\partial H} \right)\dot{H}.
\label{eq:dBGPM_dt}
\end{equation}
According to \eqref{eq:B_rev}, ${\partial B_{\scriptscriptstyle\mathrm{Rev}}}/{\partial H}=\mu'_{\scriptscriptstyle\mathrm{Rev}}(H)$. Thus, an incremental \textit{irreversible} permeability may be analogously defined~as 
\begin{equation}
\mu'_{\scriptscriptstyle\mathrm{Irr}}(H,\mathcal{A},\mathcal{B}) = \dfrac{\partial B_{\scriptscriptstyle\mathrm{Irr}}}{\partial H} = \dfrac{\hat{B}_{\scriptscriptstyle\mathrm{Irr}}}{T(\alpha_0,\beta_0)} \,\dfrac{\partial f_{\scriptscriptstyle\mathrm{CPM}}(H,\mathcal{A},\mathcal{B})}{\partial H},
\end{equation}
and therefore \eqref{eq:dBGPM_dt} may be expressed as
\begin{equation}
\dot{B} =  \mu'_{\scriptscriptstyle\mathrm{GPM}}(H,\mathcal{A},\mathcal{B}) \,\dot{H},
\label{eq:dBGPM_dt_2}
\end{equation}
where $\mu'_{\scriptscriptstyle\mathrm{GPM}}(H,\mathcal{A},\mathcal{B})$ is the incremental permeability of the GPM, which is equal to the sum of the reversible and irreversible permeabilities,
\begin{equation}
\mu'_{\scriptscriptstyle\mathrm{GPM}}(H,\mathcal{A},\mathcal{B})=\mu'_{\scriptscriptstyle\mathrm{Rev}}(H) + \mu'_{\scriptscriptstyle\mathrm{Irr}}(H,\mathcal{A},\mathcal{B}).
\end{equation}

Then, to calculate $\mu'_{\scriptscriptstyle\mathrm{Irr}}(H,\mathcal{A},\mathcal{B})$, the partial derivative of $f_{\scriptscriptstyle\mathrm{CPM}}(H,\mathcal{A},\mathcal{B})$ with respect to $H$ is obtained from \eqref{eq:yCPM_inc}--\eqref{eq:yCPM_dec} as
\begin{equation}
\dfrac{\partial f_{\scriptscriptstyle\mathrm{CPM}}(H,\mathcal{A},\mathcal{B})}{\partial H} = 
\left\{\!\!\! 
\begin{array}{r@{\hspace{1ex}}c@{\hspace{1ex}}c}
+2\,\dfrac{\partial \,T\left(H,\beta_n\right)}{\partial H}, & \text{if} & \dot{H}\geq 0 \\[1em]
-2\,\dfrac{\partial\,T\left(\alpha_n,H\right)}{\partial H}, & \text{if} & \dot{H}< 0,
\end{array}
\right.
\label{eq:partialCPM}
\end{equation}
where the partial derivatives of $T\left(H,\beta_n\right)$ and $T\left(\alpha_n,H\right)$ can be calculated from \eqref{eq:preisach_triangle} by applying Leibniz's integral rule,
\begin{align}
\dfrac{\partial \,T\left(H,\beta_n\right)}{\partial H} &= +\int_{\beta_n}^{H} P(H,\beta) \,\mathrm{d}\beta, \label{eq:partialT1}\\
\dfrac{\partial\,T\left(\alpha_n,H\right)}{\partial H} &= -\int_{H}^{\alpha_n} \! P(\alpha,H) \,\mathrm{d}\alpha. \label{eq:partialT2}
\end{align}
It must be pointed out that, when using the Cauchy-based Preisach distribution proposed in Section \ref{subsec:cpm}, these integrals can be analytically solved and, consequently, the computation of $\mu'_{\scriptscriptstyle\mathrm{GPM}}(H,\mathcal{A},\mathcal{B})$ is not numerically expensive. The resulting expressions, which may be obtained by means of any symbolic computation software, have been omitted from the paper due to space constraints.

\section{Hybrid Dynamical Model}
\label{sec:hybrid}

\begin{figure}[t]
\centering
\includegraphics[]{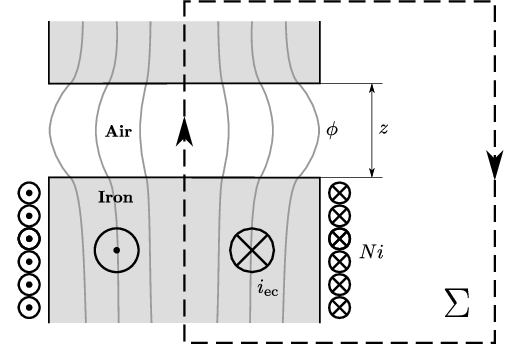}
\caption{Reluctance actuator diagram showing an air gap and part of the iron core. The arrows indicate the sign convention for $\phi$, $i$ and $i_\mathrm{ec}$.}
\label{fig:MEC}
\end{figure}

Fig.~\ref{fig:MEC} depicts the air gap and part of the core and the coil of a typical linear reluctance actuator. This diagram is used to explain our modeling methodology. The position of the mover in this actuator is defined by the length of the air gap, $z$. A coil of $N$ turns, carrying an electrical current $i$, is wrapped around the core. It generates a magnetic flux $\phi$ and induces an equivalent eddy current $i_\mathrm{ec}$ in the iron. This description is a valid representation for almost any armature arrangement, e.g., plunger-type, E-core, or C-core devices, and can also be applied to devices with rotary motion simply by using the equivalent angular variables.

The dynamical model of this actuator in state space form has three continuous state variables: an average value of the magnetic field intensity in the iron core, $H_\mathrm{iron}$, and the position and velocity of the movable armature, given by the length of the air gap, $z$, and its derivative, $\varv_z=\dot{z}$. Hence, the continuous state vector is
\begin{equation}
\mathbf{x} = \left[\, H_\mathrm{iron} \hspace{1.5ex} z \hspace{1.5ex} \varv_z \, \right]^\mathsf{T}.
\end{equation}

\subsection{Electromagnetic dynamics}

The electromagnetic part of the model is based on two equations. First, the electrical equation of the coil,
\begin{equation}
v = R\, i + N \dot{\phi},
\label{eq:electrical}
\end{equation}
where $v$ is the voltage applied across its terminals, $R$ is the internal resistance and the other variables have been previously defined. Secondly, the application of Amp\`ere's law over the surface $\Sigma$ (see Fig.~\ref{fig:MEC}), 
\begin{equation}
\oint_{\partial\Sigma} \mathbf{H} \cdot \mathrm{d} \mathbf{l} = \iint_\Sigma \mathbf{J} \cdot \mathrm{d} \mathbf{S},
\label{eq:ampere}
\end{equation}
where ${\partial\Sigma}$ is the contour of $\Sigma$, defined by the closed-loop path of the magnetic flux. The circulation of the magnetic field intensity $\mathbf{H}$ can be divided into two terms corresponding to the iron and the air. On the other hand, the surface integral of the current density $\mathbf{J}$ is equal to the sum of $N$ times the coil current plus the eddy currents. Thus, \eqref{eq:ampere} is converted into
\begin{equation}
\int_\mathrm{Iron} \mathbf{H} \cdot \mathrm{d} \mathbf{l} + \int_\mathrm{Air} \mathbf{H} \cdot \mathrm{d} \mathbf{l} = Ni + i_\mathrm{ec}.
\label{eq:ampere_2}
\end{equation}

Regarding the iron term, we take the usual assumption that the magnetic field intensity within the core can be approximated by an average value $H_\mathrm{iron}$, so that
\begin{equation}
\int_\mathrm{Iron} \mathbf{H} \cdot \mathrm{d} \mathbf{l} = H_\mathrm{iron} \, l_\mathrm{iron},
\label{eq:ampere_iron}
\end{equation}
where $l_\mathrm{iron}$ is the length of the iron core which, depending on the armature arrangement, is either a constant or a function of the length of the air gap. In addition, the magnetic flux is assumed equal to the product of an average magnetic flux density, $B_\mathrm{iron}$, and an average cross-sectional area, $A_\mathrm{iron}$, 
\begin{equation}
\phi = B_\mathrm{iron} \, A_\mathrm{iron}.
\label{eq:B_iron}
\end{equation}
The relation between $B_\mathrm{iron}$ and $H_\mathrm{iron}$ is then modeled by means of the GPM, i.e., $B_\mathrm{iron} \!=\! f_{\scriptscriptstyle\mathrm{GPM}}(H_\mathrm{iron})$. For the sake of clarity, in this part of the paper we omit the explicit dependence of $f_{\scriptscriptstyle\mathrm{GPM}}$ on the extrema sets $\mathcal{A}$ and $\mathcal{B}$.

The integral term corresponding to the air is modeled through the concept of reluctance,
\begin{equation}
\int_\mathrm{Air} \mathbf{H} \cdot \mathrm{d} \mathbf{l} = \phi \, \mathcal{R}_\mathrm{air},
\label{eq:int_air}
\end{equation}
where $\mathcal{R}_\mathrm{air}$ is the reluctance of the gap which, given that the air is a magnetically linear material, is a function only of $z$. This reluctance may be obtained by several approaches, but it is highly recommended to use a method that accurately describes the effects of flux fringing~\cite{ramirez2018reluctance}. FEM models \cite{bao2014optimization,xu2016multiphysics} and analytical expressions \cite{mclyman2016transformer} are valid options to consider.

Finally, the eddy currents are modeled assuming that the magnetic flux is uniform within the section of the iron core. In this case, it can be shown \cite{vrijsen2014magnetic} that the net induced current is proportional to the derivative of the flux,
\begin{equation}
i_\mathrm{ec}  = -k_\mathrm{ec} \,\dot{\phi},
\label{eq:i_ec}
\end{equation}
where $k_\mathrm{ec}$ depends on the geometry and the electrical conductivity of the iron core. While for most actuators it can be assumed a positive constant, in some cases it might be necessary to model its dependence on the position. As an alternative to {\eqref{eq:i_ec}}, the dynamic effects within the iron could also be modeled by means of rate-dependent Preisach models~\cite{fuzi1999computationally}.

Assuming that $v$ is the input of the model, the dynamic equation for the magnetic flux can be obtained using \eqref{eq:electrical}--\eqref{eq:i_ec},
\begin{equation}
\dot{\phi} = \frac{\dfrac{N}{R}\,v - \phi \, \mathcal{R}_\mathrm{air} - f_{\scriptscriptstyle\mathrm{GPM}}^{\,-1}\!\left({\phi}/{A_\mathrm{iron}}\right)l_\mathrm{iron}}{\dfrac{N^2}{R}+k_\mathrm{ec}},
\label{eq:dphi_dt}
\end{equation}
but, as shown, it requires the inversion of the GPM. This problem has been widely treated in the literature \cite{bobbio1997models,matsuo2003application,dlala2006inverted,hedegard2017adaptive} and the proposed solutions involve either the use of a modified version of the Preisach model that allows for explicit inversion or the numerical inversion of the forward model by means of an iterative method. As stated in \cite{dlala2006inverted}, this latter approach is particularly inefficient in terms of computation.

Nevertheless, considering that the derivatives of $B_\mathrm{iron}$ and $H_\mathrm{iron}$ are linked together by the incremental permeability of the GPM, $\dot{B}_\mathrm{iron} \!=\!  \mu'_{\scriptscriptstyle\mathrm{GPM}}(H_\mathrm{iron}) \,\dot{H}_\mathrm{iron}$, the previous set of equations can be used to solve the dynamics of the magnetic field intensity $H_\mathrm{iron}$, instead of that of $\phi$,
\begin{equation}
\dot{H}_\mathrm{iron} = \frac{\dfrac{N}{R}\,v - A_\mathrm{iron}\, f_{\scriptscriptstyle\mathrm{GPM}}(H_\mathrm{iron})\, \mathcal{R}_\mathrm{air} - H_\mathrm{iron} \, l_\mathrm{iron}}{\left(\dfrac{N^2}{R}+k_\mathrm{ec}\right)A_\mathrm{iron}\,\mu'_{\scriptscriptstyle\mathrm{GPM}}(H_\mathrm{iron})}.
\label{eq:dHiron_dt}
\end{equation}
In contrast to \eqref{eq:dphi_dt}, this expression does not require the inversion of the GPM and, consequently, the numerical solution of $H_\mathrm{iron}$  is far more efficient because it requires only one evaluation of $f_{\scriptscriptstyle\mathrm{GPM}}$ and $\mu'_{\scriptscriptstyle\mathrm{GPM}}$ per integration step. Given that these two functions depend not only on $H_\mathrm{iron}$, but also on its direction and on the sets $\mathcal{A}$ and $\mathcal{B}$, and considering that $\mathcal{R}_\mathrm{air}$ is a function of $z$, let us express \eqref{eq:dHiron_dt} symbolically as
\begin{equation}
\dot{H}_\mathrm{iron} = 
\left\{\!\!\! \begin{array}{r@{\hspace{1ex}}c@{\hspace{0.5ex}}c}
h_\mathrm{incr}(\mathbf{x},\mathcal{A},\mathcal{B},v), & \text{if} & H_\mathrm{iron} \text{ is increasing}\\[1em]
h_\mathrm{decr}(\mathbf{x},\mathcal{A},\mathcal{B},v), & \text{if} & H_\mathrm{iron} \text{ is decreasing}.
\end{array} \right.
\label{eq:f}
\end{equation}
Note that, in order to have an unambiguous definition of the direction of $H_\mathrm{iron}$, the functions $h_\mathrm{incr}$ and $h_\mathrm{decr}$ must have the same sign for any value of the inputs. Considering \eqref{eq:dHiron_dt}, this is achieved if $\mu'_{\scriptscriptstyle\mathrm{GPM}}$ is strictly positive; a condition which, on the other hand, ensures the physical meaning of the permeability. Since the irreversible term $\mu'_{\scriptscriptstyle\mathrm{Irr}}$ is always greater than or equal to zero (see Section~\ref{subsec:mu_inc}), a sufficient solution consists in fitting the reversible model \eqref{eq:mu_rev} using the constraint that $\mu'_{\scriptscriptstyle\mathrm{Rev}}$ is strictly positive for any value of $H_\mathrm{iron}$.

\subsection{Motion dynamics}

The motion of the actuator is governed by Newton's second law,
\begin{equation}
F = m \, \ddot{z} = m \, \dot{\varv}_z,
\label{eq:newton2nd}
\end{equation}
where $m$ is the mass of the movable armature and $F$ is the net force acting on this component. This force is modeled as the sum of the magnetic force produced by the actuator, the elastic force of the return spring and a damping term that accounts for friction, i.e.,
\begin{equation}
F = F_\mathrm{mag} - k_\mathrm{s}\, (z - z_\mathrm{s}) - c\, \varv_z,
\label{eq:Fnet}
\end{equation}
where $k_\mathrm{s}$ and $z_\mathrm{s}$ are respectively the spring stiffness constant and the gap length at its equilibrium position, and $c$ is the damping coefficient. Considering the concept of reluctance, the magnetic force $F_\mathrm{mag}$ can be expressed \cite{ramirez2016new} as
\begin{equation}
F_\mathrm{mag} = -\frac{1}{2}\,\phi^2 \,\frac{\partial\mathcal{R}_\mathrm{air}}{\partial z},
\label{eq:F_mag}
\end{equation}
which, given \eqref{eq:B_iron} and the relation provided by the GPM, leads to
\begin{equation}
F_\mathrm{mag} = -\frac{1}{2}\,{A_\mathrm{iron}}^{\!2}\big(f_{\scriptscriptstyle\mathrm{GPM}}(H_\mathrm{iron},\mathcal{A},\mathcal{B})\big)^2  \,\frac{\partial\mathcal{R}_\mathrm{air}}{\partial z}.
\label{eq:Fmag}
\end{equation}
Then, as shown by \eqref{eq:Fnet} and \eqref{eq:Fmag}, the net force is a function of the three continuous state variables and, given that it relies on the GPM, also of the past extrema and the direction of $H_\mathrm{iron}$. Hence, we can symbolically express it as 
\begin{equation}
F=
\left\{\!\!\! \begin{array}{r@{\hspace{1ex}}c@{\hspace{0.5ex}}c}
F_\mathrm{incr}(\mathbf{x},\mathcal{A},\mathcal{B}), & \text{if} & H_\mathrm{iron} \text{ is increasing}\\[1em]
F_\mathrm{decr}(\mathbf{x},\mathcal{A},\mathcal{B}), & \text{if} & H_\mathrm{iron} \text{ is decreasing}.
\end{array} \right.
\label{eq:F}
\end{equation}

This force drives the motion of the actuator whenever the position $z$ is between $z_\mathrm{min}$ and $z_\mathrm{max}$. These two constants represent respectively the minimum and maximum positions of the movable armature allowed by the mechanical design of the device. Since this position is described in our model by the length of the air gap, $z_\mathrm{min}$ will be equal to zero when modeling most linear reluctance actuators. 

\subsection{Hybrid Automaton}

\begin{figure*}[t!]
\scriptsize
	\centering
	\setlength{\fboxsep}{1pt}
	\setlength{\fboxrule}{0.5pt}
	\fbox{\hfill \includegraphics{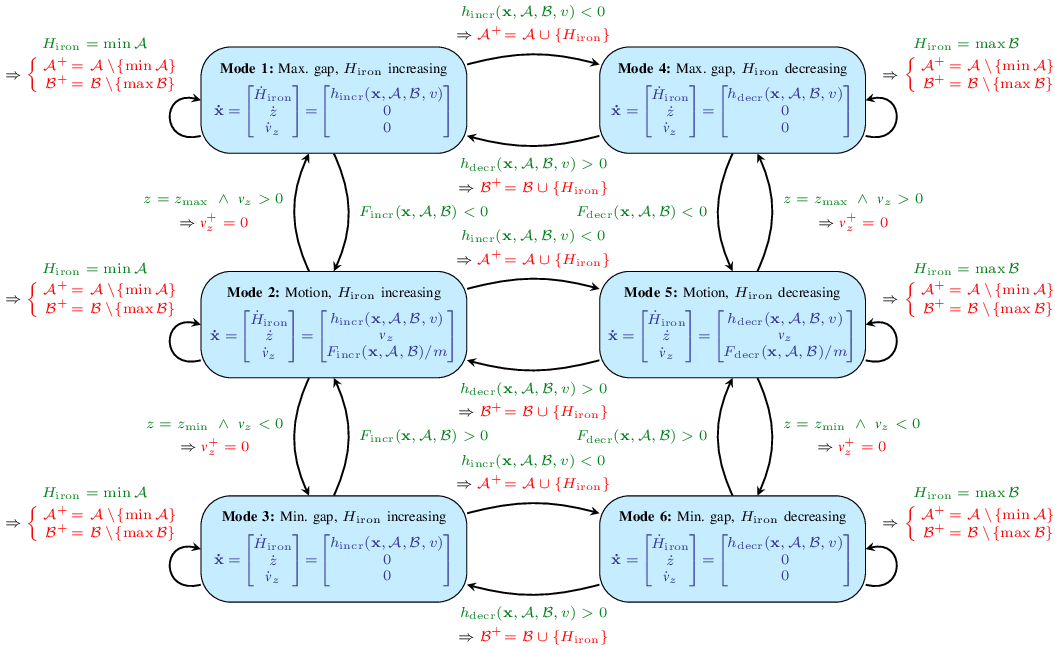} \hfill}
\caption{Hybrid automaton modeling the dynamics of the reluctance actuator.}
\label{fig:hybrid_automaton}
\end{figure*}

Considering the described motion and similarly as in \cite{moya2017nonlinear}, the position of the movable armature defines by itself three different dynamic modes: two corresponding to the two position boundaries and another one for the motion between them. These three modes are then doubled by the fact that the GPM uses different expressions depending on the direction of its input, which leads to the hybrid functions \eqref{eq:f} and \eqref{eq:F}. Consequently, the proposed hybrid dynamical model has six different dynamic modes, which correspond to:

\begin{enumerate}
\item Maximum gap and $H_\mathrm{iron}$ increasing.
\item Motion between boundaries and $H_\mathrm{iron}$ increasing.
\item Minimum gap and $H_\mathrm{iron}$ increasing.
\item Maximum gap and $H_\mathrm{iron}$ decreasing.
\item Motion between boundaries and $H_\mathrm{iron}$ decreasing.
\item Minimum gap and $H_\mathrm{iron}$ decreasing.
\end{enumerate}

The complete state of the system, $\chi$, is composed of the dynamic mode, which will be denoted by the discrete variable $q \in \left\{1,2,3,4,5,6\right\}$, the already defined continuous state $\mathbf{x}$, and, as a particular feature of our model, also the extrema sets $\mathcal{A}$ and $\mathcal{B}$. Hence, by a slight misuse of notation,
\begin{equation}
\chi = \left(q,\mathbf{x},\mathcal{A},\mathcal{B}\right).
\end{equation}

The dynamic modes are then connected by the corresponding \textit{guard} and \textit{reset} maps, leading to the hybrid automaton shown in Fig.~\ref{fig:hybrid_automaton}. In this figure, each transition is described by its guard condition (in {\color{color_guard}green}) and its reset function (in {\color{color_jump}red}), respectively before and after a right arrow ($\Rightarrow$). The superscript $+$ is used to specify the values of the continuous variables and the extrema sets after the jump. The reset function is explicitly shown only for those variables that change during the jump; if the transition does not imply a jump of $\mathbf{x}$, $\mathcal{A}$, or $\mathcal{B}$, only the guard condition is presented. Note that the continuous state $\mathbf{x}$ may flow and occasionally jump, but the dynamic mode $q$ and the extrema sets $\mathcal{A}$ and $\mathcal{B}$ change by their nature only during jumps.

Regarding the motion, the model operates as follows. If the armature is moving and reaches any of the position boundaries, the automaton jumps to the corresponding static position mode and the velocity is reset to zero. Then, when the force $F$ has the correct sign to start the motion, the automaton gets back to the motion mode. For simplicity, no bouncing phenomenon has been included in the model, but it may be easily incorporated by means of any of the well-known techniques \cite{jun2008dynamic}.

On the other hand, two types of jumps may arise due to the electromagnetic dynamics. First, those related to the wiping out property of the Preisach model. If the magnetic field varies inside a minor loop, bounded between the maximum value of the minima set, $\mathrm{max}\,\mathcal{B}$, and the minimum value of the maxima set, $\mathrm{min}\,\mathcal{A}$, and then it reaches one of the boundaries, the complete loop is wiped out from the history of the model. When this happens, the automaton jumps from its present mode to itself, and during the jump both $\mathrm{max}\,\mathcal{B}$ and $\mathrm{min}\,\mathcal{A}$ are removed from the extrema sets. The system also jumps when $H_\mathrm{iron}$ reaches an extrema. In this regard, if the magnetic field increases and then changes direction, the automaton jumps to the corresponding decreasing mode and the present value of $H_\mathrm{iron}$ is incorporated to the maxima set. Analogously, if the field first decreases and then increases, the automaton changes mode and the minima set is expanded with the new minimum. It should be noted that, since $h_\mathrm{incr}$ and $h_\mathrm{decr}$ always have the same sign, the automaton will not switch infinitely many times between the increasing and the decreasing modes in a single time instant. Thus, the Zeno phenomenon \cite{goebel2009hybrid} is prevented.

\section{Experiments and Results}
\label{sec:experiments}

A model of an actual reluctance actuator has been built to show the applicability of the proposed modeling technique. In this section, we present the device and the setup used in the experiments, explain the identification process required to determine the values of the unknown parameters and, finally, provide and discuss the main results.

\subsection{Actuator and Experimental Setup}

\begin{figure}[t]
\centering
\hfill
\begin{minipage}[b][45.5mm][c]{20mm}\centering \includegraphics[height=40mm]{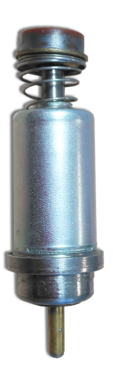} \end{minipage}
\begin{minipage}[b][45.5mm][c]{55mm}\centering \includegraphics[height=45mm]{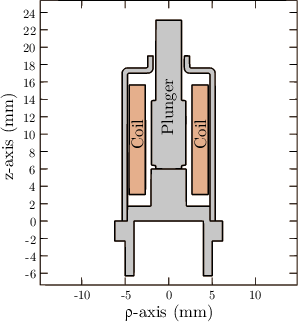} \end{minipage}
\hfill
\captionof{figure}{Solenoid valve and its geometry.}
\label{fig:valve}
\end{figure}

\begin{figure}[t]
\centering
\includegraphics[width=2.5cm]{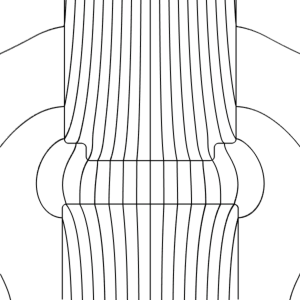}
\hfill
\includegraphics[width=2.5cm]{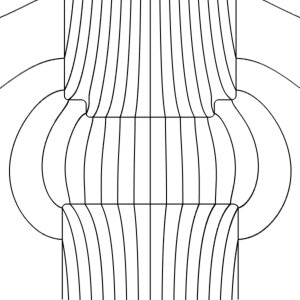}
\hfill
\includegraphics[width=2.5cm]{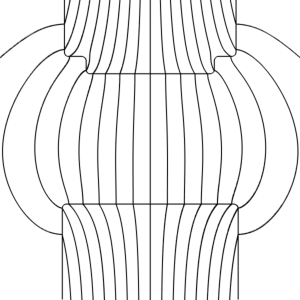}
\captionof{figure}{Magnetic flux lines in the air gap. Results from FEM simulations at gap lengths of 1, 2, and 3~mm, respectively. The FEM model is only used to characterize the air gap reluctance.}
\label{fig:valve_gap}
\end{figure}

\begin{figure}[t]
\centering
\includegraphics[]{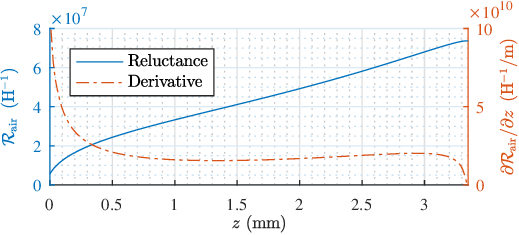}
\captionof{figure}{Air gap reluctance of the solenoid valve and partial derivative with respect to the gap length. Results from FEM simulations.}
\label{fig:reluctance_fem_2}
\end{figure}

The device utilized in the tests is a solenoid valve designed for low-pressure gas lines (see Fig.~\ref{fig:valve}). It is basically a plunger-type reluctance actuator, with the coil wrapped around the plunger and an iron housing that serves both as returning path for the flux and protective cover. A helical spring outside the magnetic circuit ensures that the plunger returns to its original position once the core is demagnetized. The material of the core is unknown and, hence, the parameters of the reversible part of the GPM ($\mu_1$, $\mu_2$, $H_1$, and $H_2$) and those corresponding to the irreversible part (${\hat{B}_{\scriptscriptstyle\mathrm{Irr}}}$, $m_{h_c}$, $s_{h_c}$, and $s_{h_m}$) as well as the eddy current parameter ($k_\mathrm{ec}$) will be determined by means of experimental tests and identification procedures. The rest of the parameters are known and can be found in Table~\ref{tab:known_parameters} (note that mechanical damping is assumed negligible). In addition, the reluctance of the air gap and its partial derivative with respect to the gap length have been obtained from an FEM model implemented in COMSOL Multiphysics (see Figs.~{\ref{fig:valve_gap}} and~{\ref{fig:reluctance_fem_2}}). In this connection, other modeling approaches can be found in~\cite{ramirez2018reluctance}. As shown, the reluctance increases rapidly for small values of the gap length and then grows approximately linearly up to the maximum gap. The reluctance value for zero gap is not zero because there exists a secondary annular gap between the movable core and the housing (see Fig.~\ref{fig:valve}).

\begin{table}[t]
\renewcommand{\arraystretch}{1.25}
\captionof{table}{Model parameters}
\label{tab:known_parameters}
\centering
	\begin{tabular}{ccc}
		\hline
		{\bfseries Parameter} & {\bfseries Value}\\
		\hline
		$R$ & { 49 $\Omega$}\\
		$N$ & { 1200}\\
		$l_\mathrm{iron}$ & { 55 mm}\\
		$A_\mathrm{iron}$ & { 12.57 mm$^2$}\\
		\hline
	\end{tabular}
	\hspace{1em}
	\begin{tabular}{ccc}
		\hline
		{\bfseries Parameter} & {\bfseries Value}\\
		\hline
		$m$ & { 1.6 g}\\
		$k_\mathrm{s}$ & { 55 N/m}\\
		$z_\mathrm{s}$ & { 15 mm}\\
		$c$ & { 0 Ns/m}\\
		\hline
	\end{tabular}
\renewcommand{\arraystretch}{1.0}
\end{table}

\begin{figure}[t]
\centering
\includegraphics[]{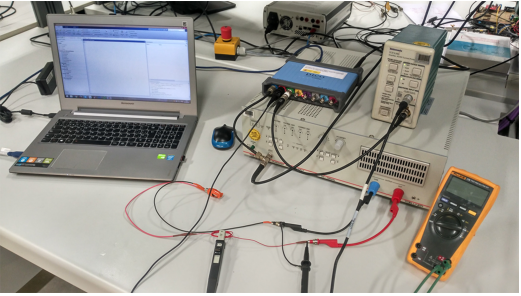}
\captionof{figure}{Experimental setup used during the tests.}
\label{fig:testbench}
\end{figure}

The experimental setup used in the tests is shown in Fig.~\ref{fig:testbench} and consists of the following equipment: a Toellner TOE~7621 four-quadrant voltage and current amplifier, a Picoscope~4824 USB oscilloscope that also features an arbitrary waveform generator, a Tektronix TCP312A current probe and its corresponding TCPA300 amplifier, a personal computer with Matlab and the Instrument Control Toolbox and, finally, the valve, which can also be seen in the picture, wrapped in orange electrical tape. The operation of the setup is as follows. First, a current or voltage waveform is designed and programmed in Matlab. When the code is executed, the reference signal is sent to the generator, amplified by the power supply and applied to the valve. At the same time, the oscilloscope measures, registers, and sends the values of voltage, $v$, and current, $i$,  back to the computer. Finally, the data are processed and the magnetic flux, $\phi$, is estimated as described in \cite{ramirez2018real}.

Three different sets of experiments have been performed, each having a specific purpose and based on a particular type of waveform which is repeated at several levels of excitation. The first group of tests is used to determine the values of the parameters of the GPM and have been carried out with the plunger fixed at zero gap ($z=0$). These use a low-frequency (10 Hz) sinusoidal current input, which minimizes the induced currents and, consequently, improves the identifiability of the hysteresis model. Once the GPM parameters are determined, the second group of experiments is utilized to identify the value of~$k_\mathrm{ec}$. In these tests the position is also externally set at $z=0$, but, instead of the sinusoidal current, a symmetric square wave voltage is used to induce transient eddy currents in the iron core. Finally, the third set, which is only for validation purposes, consists of a series of normal operating cycles of the valve at different voltages.
That is, the plunger is released, the valve is inserted into its corresponding gas faucet, which limits the motion of the plunger from $z_\mathrm{min}\!=\!0$~mm up to $z_\mathrm{max}\!=\!0.9$~mm, and finally the device is activated using a unipolar square wave voltage. Table~\ref{tab:tests_sets} sums up the described sets of experiments. It must be noted that, in order to start the tests from a known initial state, a decreasing sine wave current has been applied to the valve before each group of experiments to obtain a demagnetized state in the core (degaussing process).

\begin{table}[t]
\renewcommand{\arraystretch}{1.25}
\captionof{table}{Sets of experiments}
\label{tab:tests_sets}
\centering
	\begin{tabular}{cccc}
		\hline
		{\bfseries Set} & {\bfseries Input wave} & {\bfseries Levels} & {\bfseries Purpose}\\
		\hline
		1 & Current, sinusoidal & 8 (25-500 mA) & GPM identification\\
		2 & Voltage, bipolar square & 8 (1-18 V) & $k_\mathrm{ec}$ identification\\
		3 & Voltage, unipolar square & 5 (18-26 V) & Validation\\
		\hline
	\end{tabular}
\renewcommand{\arraystretch}{1.0}
\end{table}

\subsection{Parameter Identification}

\begin{figure}[tp]
\centering
\includegraphics[]{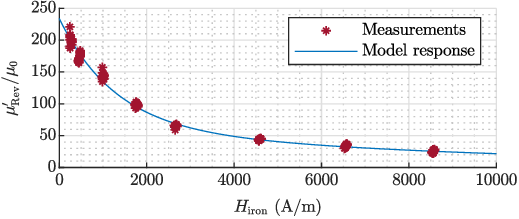}
\caption{Measured values of the reversible part of the incremental permeability and least-squares fit to equation {\eqref{eq:mu_rev}}.}
\label{fig:incremental_permeability}
\vspace{\floatsep}
\includegraphics[]{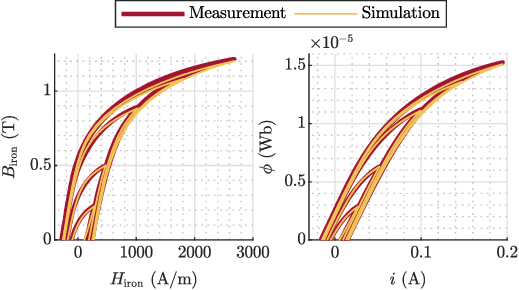}
\caption{Measured hysteresis curves and simulated output of the GPM.}
\label{fig:fitting_GPM}
\vspace{\floatsep}
\includegraphics[]{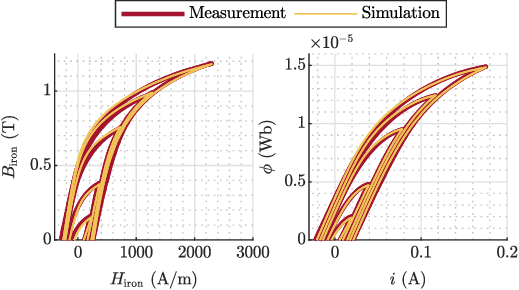}
\caption{Measured response of the system under the presence of hysteresis and eddy currents and simulated output of the hybrid dynamical model.}
\label{fig:fitting_ec}
\end{figure}

The identification process of the GPM is similar to that presented in \cite{vrijsen2014magnetic} and is carried out using the data from the first set of tests. Considering the measured current $i$ and  estimated flux $\phi$, the magnetic flux density is first obtained from \eqref{eq:B_iron} as \mbox{$B_\mathrm{iron}\! =\! \phi / A_\mathrm{iron}$}. On the other hand, the magnetic field intensity is calculated using \eqref{eq:ampere_2}, \eqref{eq:ampere_iron} and \eqref{eq:int_air}, and assuming that the induced currents are negligible, i.e.,
\begin{equation}
H_\mathrm{iron} = \big(Ni - \phi\, \mathcal{R}_\mathrm{air}(z=0) \big) / \, l_\mathrm{iron}.
\label{eq:Hiron_no_ec}
\end{equation}
The reversible part of the model is identified firstly. For this purpose, measurements of $\mu'_\mathrm{Rev}$ are obtained as the slope $\partial B_\mathrm{iron}/\partial H_\mathrm{iron}$ at the reversal points of 64 minor loops at eight different excitation levels. Then, the resulting points are used to fit the model proposed for the reversal permeability, \eqref{eq:mu_rev}, by minimizing the root-mean-square error (RMSE). The measured points and the fitted model response are shown in Fig.~\ref{fig:incremental_permeability}.

Once the parameters of the reversible part are determined, the rest of the GPM is identified. In order to match the demagnetized state of the valve, the model also has to be first demagnetized, which is simply achieved by setting proper initial values for the sets $\mathcal{A}$ and $\mathcal{B}$. The selection of the number of elements in these sets is a trade-off between model accuracy and computational time. In this case, we use two symmetric sets composed of a hundred elements in the interval $[-10^4,\ +10^4]$~A/m, which is the expected range for the magnetic field $H_\mathrm{iron}$.
\begin{align}
\mathcal{A}(0) &= \left\{\, \alpha_k = +10^4-100k \ \text{A/m} \mid k = 0,1,...,99 \,\right\} \\
\mathcal{B}(0) &= \left\{\, \beta_k  = -10^4+100k \ \text{A/m} \mid k = 0,1,...,99 \,\right\}
\end{align}
As explained in Section~\ref{subsec:cpm} and given that $\left|\mathcal{B}(0)\right| = \left|\mathcal{A}(0)\right|$, $H_\mathrm{iron}$ is assumed to increase at $t=0$. The GPM, \eqref{eq:BGPM}, is then computed using as input the experimental values of $H_\mathrm{iron}$ obtained from \eqref{eq:Hiron_no_ec}, and fitted by minimizing the RMSE between the simulated and the measured values of $B_\mathrm{iron}$. Fig.~\ref{fig:fitting_GPM} shows the measured hysteresis curves and the simulated response of the GPM after the identification process, in both the $B_\mathrm{iron}$\,--\,$H_\mathrm{iron}$ and $\phi$\,--\,$i$ planes. Note that, although eight different excitation levels have been used to fit the model, only the smallest five cycles are represented for clarity reasons. All the identified parameters of the GPM are shown in Table~\ref{tab:gpm_parameters}.

\begin{table}[tp]
\renewcommand{\arraystretch}{1.25}
\caption{Identified GPM parameters}
\label{tab:gpm_parameters}
\centering
	\begin{tabular}{ccc}
		\hline
		{\bfseries Parameter} & {\bfseries Value}\\
		\hline
		$\mu_1$ & { 168.8 $\mu_0$}\\
		$\mu_2$ & { 64.13 $\mu_0$}\\
		$H_1$ & { 1262 A/m}\\
		$H_2$ & { 8821 A/m}\\
		\hline
	\end{tabular}
	\hspace{1em}
	\begin{tabular}{ccc}
		\hline
		{\bfseries Parameter} & {\bfseries Value}\\
		\hline
		${\hat{B}_{\scriptscriptstyle\mathrm{Irr}}}$ & { 0.8103 T}\\
		$m_{h_c}$ & { 227.9 A/m}\\
		$s_{h_c}$ & { 154.9 A/m}\\
		$s_{h_m}$ & { 138.0 A/m}\\
		\hline
	\end{tabular}
\renewcommand{\arraystretch}{1.0}
\end{table}

The second group of tests is subsequently used to determine the value of the eddy current parameter, $k_\mathrm{ec}$, which is assumed constant. During this stage of the identification, the complete hybrid dynamical model is utilized instead of only the GPM. However, since the position of the plunger is externally set to $z\!=\!z_\mathrm{min}\!=\!0$~mm, only the dynamic modes 3 and 6 are active. The aim of the identification is to minimize the following weighted RMSE,
\begin{equation}
\sqrt{ \frac{\sum_k \left(i_{k,\mathrm{exp}} - i_{k,\mathrm{sim}}\right)^2}{\sum_k {i_{k,\mathrm{exp}}} ^2} + \frac{\sum_k \left(\phi_{k,\mathrm{exp}} - \phi_{k,\mathrm{sim}}\right)^2}{\sum_k {\phi_{k,\mathrm{exp}}}^2}},
\end{equation}
where the subscript $k$ denotes the time step and ``exp'' and ``sim'' refer respectively to the values measured in the experiments and those obtained by simulating the model. The result of the optimization leads to a value of $k_\mathrm{ec}$ equal to 1637 A/V. The responses of both the system and the model under the presence of hysteresis and eddy currents are shown in Fig.~\ref{fig:fitting_ec}. Note that, while the $B_\mathrm{iron}$\,--\,$H_\mathrm{iron}$ relation remains unchanged because it is only affected by hysteresis and saturation, the $\phi$\,--\,$i$ cycle becomes wider due to the effect of eddy currents. 

\subsection{Model Validation and Analysis}

\begin{figure}[t]
\centering
\includegraphics[]{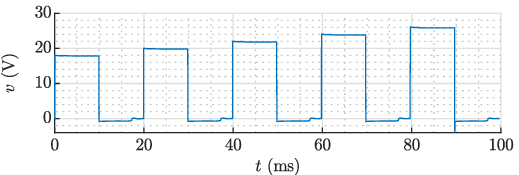}
\vspace{0.25\floatsep}
\includegraphics[]{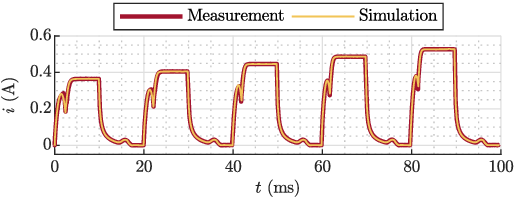}
\vspace{0.25\floatsep}
\includegraphics[]{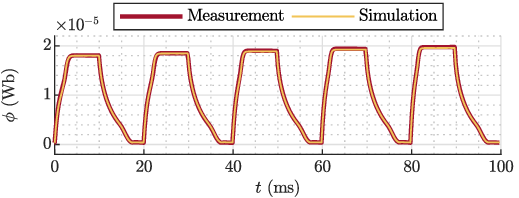}
\vspace{0.25\floatsep}
\includegraphics[]{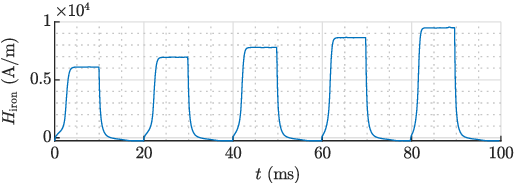}
\vspace{0.25\floatsep}
\includegraphics[]{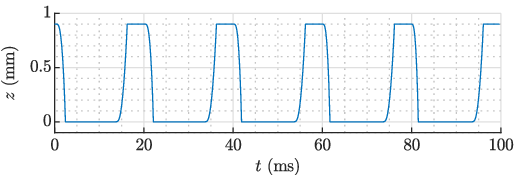}
\vspace{0.25\floatsep}
\includegraphics[]{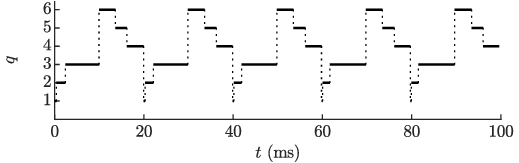}
\caption{Validation of the hybrid dynamical model. From top to bottom, voltage (input), measured and simulated current, measured and simulated flux, simulated magnetic field intensity, simulated position, and simulated dynamic mode.}
\label{fig:validation}
\end{figure}

The performance of the hybrid dynamical model is evaluated using the data of the third set of experiments. It should be recalled that, unlike in the other two sets of tests, in this case the plunger is allowed to move freely between $z_\mathrm{min}\!=\!0$~mm and $z_\mathrm{max}\!=\!0.9$~mm. Thus, all the dynamic modes of the hybrid automaton (see Fig.~\ref{fig:hybrid_automaton}) may be reached. The results of the validation process are shown in Fig.~\ref{fig:validation}. Five different square wave voltage pulses (see the first plot of the figure) have been used as input of both the real actuator and the dynamical model. The second and third plots show the measured and simulated values of current and flux. It can be seen that the dynamic response of the model matches very well with the measurements. In this regard, the RMSE of the current is 5.337~mA, which represents about 2.46$\%$ of the mean value of $i$, and the RMSE of the flux is 1.225$\cdot 10^{-7}$~Wb, which is about 1.13$\%$ of the mean value of $\phi$. Therefore, it can be concluded that the predictions given by the model are very accurate, even though the presented 100 millisecond simulation takes less than 4 seconds of computation on a fourth generation Intel~i7 processor.

Additionally, Fig.~\ref{fig:validation} also represents the simulated values of the magnetic field intensity within the iron, the position of the plunger and the dynamic mode of the automaton. As shown, the behavior of the system during the five pulses is qualitatively similar. The plunger is initially at rest at $z\!=\!z_\mathrm{max}$ (dynamic mode 1) and, once the voltage pulse is applied, both the current and the flux begin to increase rapidly. After some time, the magnetic flux reaches a value such that the magnetic force given by \eqref{eq:F_mag} exceeds the spring force and, consequently, the plunger starts moving ($q\!=\!2$). Although depending on the applied voltage, the motion is always very fast and the minimum gap position is reached within the next 2 ms. Since mechanical bouncing has not been modeled, the position remains static ($q\!=\!3$) while the electromagnetic variables evolve towards a stationary state. Then, when the voltage is cut off, $H_\mathrm{iron}$ starts to decrease and the automaton jumps to dynamic mode 4. The system remains at $z\!=\!z_\mathrm{min}$ until the magnetic force is not large enough to maintain the position. At that moment, the model jumps to $q\!=\!5$ and the plunger starts the backward motion, which lasts about 2.6~ms independently of the excitation level. Finally, the position reaches $z\!=\!z_\mathrm{min}$, the automaton jumps to $q\!=\!6$, and the flux and the current return again to the initial state.

\section{Conclusions}
\label{sec:conclusions}

This paper presents a novel dynamical model for short-stroke reluctance actuators that considers magnetic hysteresis and saturation, eddy currents, flux fringing and the motion of the armature. This model, which is computationally inexpensive compared to FEM simulations, relies on an explicit dynamic solution of the generalized Preisach model which has been derived in the paper by means of the concept of incremental permeability. In this regard, it is shown that the selection of the magnetic field intensity as a state variable, instead of the magnetic flux, is the key to obtain a solution which does not require the inversion of the GPM.

The structure of the Preisach model, whose output depends on the direction of the input, together with the fact that the motion of the armature is limited by both upper and lower bounds, causes the system to exhibit hybrid behavior. Consequently, the actuator is modeled by means of a hybrid automaton with six different dynamic modes. An innovative feature of the model is that, in addition to the continuous variables (magnetic field, position, and velocity) and the dynamic mode, it also has two sets of scalars as state variables. These sets, which collect the previous maxima and minima of the magnetic field, are inputs of the GPM and evolve at discrete instants when past extrema are wiped out or new extrema are encountered.

The modeling methodology has been illustrated using a solenoid valve as an application example. The identification procedure has been thoroughly described so that it can be used to dynamically model any other reluctance actuator. The validation experiments have shown that the model is fast and provides very accurate predictions, which makes it particularly well suited for designing estimation and control algorithms. The parametric nature of the model can be exploited, e.g., to improve the design of an actuator or to perform sensitivity analyses of the parameters.

\section*{ACKNOWLEDGMENT}

E. Ramirez-Laboreo would like to express his gratitude to Professor E. A. Lomonova and all the staff of the EPE group for their warm welcome and professional support during his three-month research stay at the Eindhoven University of Technology. E.~Ramirez-Laboreo and C.~Sagues want also to thank S. Llorente for his support and BSH Home Appliances Spain for supplying part of the equipment used in this research.

\bibliographystyle{IEEEtran}

\begin{thebibliography}{10}
\providecommand{\url}[1]{#1}
\csname url@samestyle\endcsname
\providecommand{\newblock}{\relax}
\providecommand{\bibinfo}[2]{#2}
\providecommand{\BIBentrySTDinterwordspacing}{\spaceskip=0pt\relax}
\providecommand{\BIBentryALTinterwordstretchfactor}{4}
\providecommand{\BIBentryALTinterwordspacing}{\spaceskip=\fontdimen2\font plus
\BIBentryALTinterwordstretchfactor\fontdimen3\font minus \fontdimen4\font\relax}
\providecommand{\BIBforeignlanguage}[2]{{%
\expandafter\ifx\csname l@#1\endcsname\relax
\typeout{** WARNING: IEEEtran.bst: No hyphenation pattern has been}%
\typeout{** loaded for the language `#1'. Using the pattern for}%
\typeout{** the default language instead.}%
\else
\language=\csname l@#1\endcsname
\fi
#2}}
\providecommand{\BIBdecl}{\relax}
\BIBdecl

\bibitem{katalenic2016high}
A.~Katalenic, H.~Butler, and P.~P.~J. van~den Bosch, ``High-precision force control of short-stroke reluctance actuators with an air gap observer,'' \emph{IEEE/ASME Trans. Mechatronics}, vol.~21, no.~5, pp. 2431--2439, 2016.

\bibitem{bao2014optimization}
J.~Bao, N.~H. Vrijsen, B.~L.~J. Gysen, R.~L.~J. Sprangers, and E.~A. Lomonova, ``Optimization of the force density for medium-stroke reluctance actuators,'' \emph{IEEE Trans. Ind. Appl.}, vol.~50, no.~5, pp. 3194--3202, Sep. 2014.

\bibitem{enrici2016design}
P.~Enrici, F.~Dumas, N.~Ziegler, and D.~Matt, ``Design of a high-performance multi-air gap linear actuator for aeronautical applications,'' \emph{IEEE Trans. Energy Convers.}, vol.~31, no.~3, pp. 896--905, Sep. 2016.

\bibitem{kimiabeigi2016high}
M.~Kimiabeigi, J.~Widmer, R.~Long, Y.~Gao, J.~Goss, R.~Martin, T.~Lisle, J.~S. Vizan, A.~Michaelides, and B.~Mecrow, ``High-performance low-cost electric motor for electric vehicles using ferrite magnets,'' \emph{IEEE Trans. Ind. Electron.}, vol.~63, no.~1, pp. 113--122, Jan. 2016.

\bibitem{mercorelli2012antisaturating}
P.~Mercorelli, ``An antisaturating adaptive preaction and a slide surface to achieve soft landing control for electromagnetic actuators,'' \emph{IEEE/ASME Trans. Mechatronics}, vol.~17, no.~1, pp. 76--85, Feb. 2012.

\bibitem{yang2013multiobjective}
Y.~P. Yang, J.~J. Liu, D.~H. Ye, Y.~R. Chen, and P.~H. Lu, ``Multiobjective optimal design and soft landing control of an electromagnetic valve actuator for a camless engine,'' \emph{IEEE/ASME Trans. Mechatronics}, vol.~18, no.~3, pp. 963--972, Jun. 2013.

\bibitem{ramirez2017new}
E.~Ramirez-Laboreo, C.~Sagues, and S.~Llorente, ``A new run-to-run approach for reducing contact bounce in electromagnetic switches,'' \emph{IEEE Trans. Ind. Electron.}, vol.~64, no.~1, pp. 535--543, Jan. 2017.

\bibitem{Lee2015control}
J.-H. Lee, Y.-W. Yun, H.-W. Hong, and M.-K. Park, ``Control of spool position of on/off solenoid operated hydraulic valve by sliding-mode controller,'' \emph{J. Mech. Sci. Technol.}, vol.~29, no.~12, pp. 5395--5408, Dec. 2015.

\bibitem{Zhao2016linear}
X.~Zhao, L.~Li, J.~Song, C.~Li, and X.~Gao, ``Linear control of switching valve in vehicle hydraulic control unit based on sensorless solenoid position estimation,'' \emph{IEEE Trans. Ind. Electron.}, vol.~63, no.~7, pp. 4073--4085, Jul. 2016.

\bibitem{STRAUBERGER2016206}
F.~Strau{\ss}berger and J.~Reuter, ``Position estimation in electro-magnetic actuators taking into account hysteresis effects,'' \emph{IFAC-PapersOnLine}, vol.~49, no.~21, pp. 206 -- 212, 2016.

\bibitem{vrijsen2014prediction}
N.~H. Vrijsen, J.~W. Jansen, and E.~A. Lomonova, ``Prediction of magnetic hysteresis in the force of a prebiased e-core reluctance actuator,'' \emph{IEEE Trans. Ind. Appl.}, vol.~50, no.~4, pp. 2476--2484, Jul. 2014.

\bibitem{fang2016novel}
S.~Fang, Q.~Liu, H.~Lin, and S.~L. Ho, ``A novel flux-weakening control strategy for permanent-magnet actuator of vacuum circuit breaker,'' \emph{IEEE Trans. Ind. Electron.}, vol.~63, no.~4, pp. 2275--2283, Apr. 2016.

\bibitem{ramirez2016new}
E.~Ramirez-Laboreo, C.~Sagues, and S.~Llorente, ``A new model of electromechanical relays for predicting the motion and electromagnetic dynamics,'' \emph{IEEE Trans. Ind. Appl.}, vol.~52, no.~3, pp. 2545--2553, May/Jun. 2016.

\bibitem{xu2016multiphysics}
B.~Xu, R.~Ding, J.~Zhang, L.~Sha, and M.~Cheng, ``Multiphysics-coupled modeling: Simulation of the hydraulic-operating mechanism for a sf6 high-voltage circuit breaker,'' \emph{IEEE/ASME Trans. Mechatronics}, vol.~21, no.~1, pp. 379--393, Feb. 2016.

\bibitem{Gaeta2013experimental}
A.~di~Gaeta, U.~Montanaro, and V.~Giglio, ``Experimental validation of a hybrid analytical-fem model of an electromagnetic engine valve actuator and its control application,'' \emph{IEEE/ASME Trans. Mechatronics}, vol.~18, no.~2, pp. 807--812, Apr. 2013.

\bibitem{guofo2011output}
Z.~Guofo, W.~Qiya, and R.~Wanbin, ``An output space-mapping algorithm to optimize the dimensional parameter of electromagnetic relay,'' \emph{IEEE Trans. Magn.}, vol.~47, no.~9, pp. 2194--2199, Sep. 2011.

\bibitem{wattiaux2011modelling}
D.~Wattiaux and O.~Verlinden, ``Modelling of the dynamic behaviour of electromechanical relays for the analysis of sensitivity to shocks and vibrations,'' \emph{Exp. Mech.}, vol.~51, no.~9, pp. 1459--1472, Nov. 2011.

\bibitem{gysen2010general}
B.~L.~J. Gysen, K.~J. Meessen, J.~J.~H. Paulides, and E.~A. Lomonova, ``General formulation of the electromagnetic field distribution in machines and devices using fourier analysis,'' \emph{IEEE Trans. Magn.}, vol.~46, no.~1, pp. 39--52, Jan. 2010.

\bibitem{jun2008dynamic}
X.~Jun, H.~Jun-jia, and Z.~Chun-yan, ``A dynamic model of electromagnetic relay including contact bounce,'' in \emph{2008 Int. Conf. Elect. Mach. and Syst.}\hskip 1em plus 0.5em minus 0.4em\relax IEEE, Oct. 2008, pp. 4144--4149.

\bibitem{lin2013design}
H.~Lin, X.~Wang, S.~Fang, P.~Jin, and S.~Ho, ``Design, optimization, and intelligent control of permanent-magnet contactor,'' \emph{IEEE Trans. Ind. Electron.}, vol.~60, no.~11, pp. 5148--5159, Nov. 2013.

\bibitem{BRAUN2017782}
T.~Braun, J.~Reuter, and J.~Rudolph, ``A novel observer approach for self sensing of single-coil digital valves,'' \emph{IFAC-PapersOnLine}, vol.~50, no.~1, pp. 782 -- 787, 2017, 20th IFAC World Congr.

\bibitem{mackenzie2016real}
I.~MacKenzie and D.~L. Trumper, ``Real-time hysteresis modeling of a reluctance actuator using a sheared-hysteresis-model observer,'' \emph{IEEE/ASME Trans. Mechatronics}, vol.~21, no.~1, pp. 4--16, Feb. 2016.

\bibitem{vrijsen2014magnetic}
N.~Vrijsen, ``Magnetic hysteresis phenomena in electromagnetic actuation systems,'' Ph.D. dissertation, Eindhoven Univ. of Technology, 2014.

\bibitem{goebel2009hybrid}
R.~Goebel, R.~G. Sanfelice, and A.~R. Teel, ``Hybrid dynamical systems,'' \emph{IEEE Control Syst. Mag.}, vol.~29, no.~2, pp. 28--93, Apr. 2009.

\bibitem{preisach1935magnetische}
F.~Preisach, ``{\"U}ber die magnetische nachwirkung,'' \emph{Zeitschrift f{\"u}r physik}, vol.~94, no. 5-6, pp. 277--302, May 1935.

\bibitem{mayergoyz1986mathematical}
I.~Mayergoyz, ``Mathematical models of hysteresis,'' \emph{IEEE Trans. Magn.}, vol.~22, no.~5, pp. 603--608, Sep. 1986.

\bibitem{rouve1995application}
L.-L. Rouve, T.~Waeckerle, and A.~Kedous-Lebouc, ``Application of preisach model to grain oriented steels: comparison of different characterizations for the preisach function p($\alpha$,$\beta$),'' \emph{IEEE Trans. Magn.}, vol.~31, no.~6, pp. 3557--3559, Nov. 1995.

\bibitem{azzerboni2003remarks}
B.~Azzerboni, E.~Cardelli, G.~Finocchio, and F.~La~Foresta, ``Remarks about preisach function approximation using lorentzian function and its identification for nonoriented steels,'' \emph{IEEE Trans. Magn.}, vol.~39, no.~5, pp. 3028--3030, Sep. 2003.

\bibitem{pruksanubal2006determination}
P.~Pruksanubal, A.~Binner, and K.~H. Gonschorek, ``Determination of distribution functions and parameters for the preisach hysteresis model,'' in \emph{17th Int. Zurich Symp. Electromagnetic Compatibility}.\hskip 1em plus 0.5em minus 0.4em\relax IEEE, Feb. 2006, pp. 258--261.

\bibitem{mayergoyz1988generalized}
I.~Mayergoyz and G.~Friedman, ``Generalized preisach model of hysteresis,'' \emph{IEEE Trans. Magn.}, vol.~24, no.~1, pp. 212--217, Jan. 1988.

\bibitem{ramirez2018reluctance}
E.~Ramirez-Laboreo and C.~Sagues, ``Reluctance actuator characterization via fem simulations and experimental tests,'' \emph{Mechatronics}, vol.~56, pp. 58 -- 66, Dec. 2018.

\bibitem{mclyman2016transformer}
C.~W.~T. McLyman, \emph{Transformer and inductor design handbook}.\hskip 1em plus 0.5em minus 0.4em\relax CRC press, 2016.

\bibitem{fuzi1999computationally}
J.~F{\"u}zi, ``Computationally efficient rate dependent hysteresis model,'' \emph{COMPEL-Int. J. Computation and Math. Elect. and Electron. Eng.}, vol.~18, no.~3, pp. 445--457, 1999.

\bibitem{bobbio1997models}
S.~Bobbio, G.~Milano, C.~Serpico, and C.~Visone, ``Models of magnetic hysteresis based on play and stop hysterons,'' \emph{IEEE Trans. Magn.}, vol.~33, no.~6, pp. 4417--4426, Nov. 1997.

\bibitem{matsuo2003application}
T.~Matsuo, D.~Shimode, Y.~Terada, and M.~Shimasaki, ``Application of stop and play models to the representation of magnetic characteristics of silicon steel sheet,'' \emph{IEEE Trans. Magn.}, vol.~39, no.~3, pp. 1361--1364, May 2003.

\bibitem{dlala2006inverted}
E.~Dlala, J.~Saitz, and A.~Arkkio, ``Inverted and forward preisach models for numerical analysis of electromagnetic field problems,'' \emph{IEEE Trans. Magn.}, vol.~42, no.~8, pp. 1963--1973, 2006.

\bibitem{hedegard2017adaptive}
M.~Hedeg{\"a}rd, T.~Wik, and C.~Wallin, ``Adaptive hysteresis compensation using reduced memory sequences,'' \emph{IEEE/ASME Trans. Mechatronics}, vol.~22, no.~5, pp. 2296--2307, Jun. 2017.

\bibitem{moya2017nonlinear}
E.~Moya-Lasheras, C.~Sagues, E.~Ramirez-Laboreo, and S.~Llorente, ``Nonlinear bounded state estimation for sensorless control of an electromagnetic device,'' in \emph{IEEE Conf. Decision and Control}.\hskip 1em plus 0.5em minus 0.4em\relax IEEE, Dec. 2017, pp. 5050--5055.

\bibitem{ramirez2018real}
E.~Ramirez-Laboreo, E.~Moya-Lasheras, and C.~Sagues, ``Real-time electromagnetic estimation for reluctance actuators,'' \emph{IEEE Trans. Ind. Electron.}, vol.~66, no.~3, pp. 1952--1961, Mar. 2018.

\end{thebibliography}

\end{document}